# GRAPHICAL MODELS FOR ZERO-INFLATED SINGLE CELL GENE EXPRESSION

By Andrew McDavid* Raphael Gottardo†,‡ Noah Simon§ and Mathias Drton‡

*Department of Biostatistics and Computational Biology*, University of Rochester Medical Center; Rochester, New York. Vaccine and Infectuous Disease Division†, Fred Hutchinson Cancer Research Center, Department of Statistics‡ and Department of Biostatistics§, University of Washington; Seattle, Washington.*

Bulk gene expression experiments relied on aggregations of thousands of cells to measure the average expression in an organism. Advances in microfluidic and droplet sequencing now permit expression profiling in single cells. This study of cell-to-cell variation reveals that individual cells lack detectable expression of transcripts that appear abundant on a population level, giving rise to zero-inflated expression patterns. To infer gene co-regulatory networks from such data, we propose a multivariate Hurdle model. It is comprised of a mixture of singular Gaussian distributions. We employ neighborhood selection with the pseudo-likelihood and a group lasso penalty to select and fit undirected graphical models that capture conditional independences between genes. The proposed method is more sensitive than existing approaches in simulations, even under departures from our Hurdle model. The method is applied to data for T follicular helper cells, and a high-dimensional profile of mouse dendritic cells. It infers network structure not revealed by other methods; or in bulk data sets. An R implementation is available at `https://github.com/amcdavid/HurdleNormal`.

**1. Introduction.** Graphical models have been used to synthesize high-throughput gene expression experiments into understandable, canonical forms [Dobra et al., 2004, Markowetz and Spang, 2007]. Although inferring causal relationships between genes is perhaps the ultimate goal of such analysis, causal models may be difficult to estimate with observational data, and experimental manipulation of specific genes has remained costly, and largely inimitable to high-throughput biology. Many analyses have thus focused on undirected graphical models (also known as Markov random fields) that capture the conditional independences present between gene expression levels. The graph determining such a model describes each gene's statistical predictors: each gene is optimally predicted using only its neighbors in the graph. With gene expression studies serving as key motivation, a host of dif-ferent approaches have been developed for structure learning and parameter estimation in undirected graphical models [Drton and Maathuis, 2017].

Characterization of the conditional independences between genes answers a variety of scientific questions. It can help falsify models of gene regulation, since statistical dependence is expected, given causal dependence. In immunology, polyfunctional immune cells, which simultaneously and non-independently express multiple cytokines, are useful predictors of vaccine response [Precopio et al., 2007]. Simultaneous expression or *co-expression* of cellular surface markers potentially define new cellular phenotypes [Lin et al., 2015], so expanding the "dictionary" of co-expression allows phenotypic refinements. Graphical models allow one to study such co-expression at the level of direct interactions.

1.1. *Single cell gene expression.* Established technology determines gene expression levels by assaying bulk aggregates of cells assayed through microarrays or RNA sequencing. Although graphical modeling of the resulting data has seen profitable applications, see e.g. Li et al. [2015], there is an inheritant limitation to what can be inferred from expression levels that are averages across hundreds or thousands of individual cells, as we discuss in Section 2. In contrast, recent microfluidic and molecular barcoding advances have enabled the measurement of the minute quantities of mRNA present in *single cells*. This new technology provides a unique resolution of gene co-expression and has the potential to facilitate more interpretable conclusions from multivariate data analysis and, in particular, graphical modeling.

At the same time, single cell expression experiments bring about new statistical challenges. Indeed, a distinctive feature of single cell gene expression data—across methods and platforms—is the bimodality of expression values [Finak et al., 2015, Marinov et al., 2014, Shalek et al., 2014]. Genes can be 'on', in which case a positive expression measure is recorded, or they can be 'off', in which case the recorded expression is zero or negligible. Although the cause of this *zero-inflation* remains unresolved, its properties are of intrinsic interest [Kim and Marioni, 2013]. It has been argued that the zero-inflation represents censoring of expression below a substantial limit of detection, yet comparison of *in silico* signal summation from many single cells, to the signal measured in biological sums of cells suggest that the limit of detection is negible [McDavid et al., 2013]. Moreover, the empirical distribution of the log-transformed counts appears rather different than would be expected from censoring: the distribution of the log-transformed, positive values is generally symmetric. Yet the presence of bimodality in technically replicated experiments ("Pool/split" experiments) implicates the involvement of technical factors [Marinov et al., 2014].

Zero-inflation is seen, in particular, in a single cell gene expression experiment we analyze in Section 6. The experiment concerns T follicular helper (Tfh) cells, which are a class of $CD4^+$ lymphocytes. B-cells that secrete antibodies require Tfh cell co-stimulation to become active [Ma et al., 2012]. Tfh cells are defined, and identified both through their location in the B-cell germinal centers, as well as their production of high levels of the proteins CXCR5, PD1 and BcL-6. In the experiment we consider, Tfh cells were identified from $CD4^+CXCR5^+PD1^+$ cells from lymph node biopsy. Figure 1 shows the pairwise expression distribution of four Tfh marker genes ($P < 10^{-20}$ compared to non-Tfh lymph node T-cells, which are not shown). Although the expression of these genes could help discriminant Tfh from non-Tfh cells, the strength of linear relationships within Tfh cells (upper panels) varies. To identify coexpressing subsets of cells or to clarify the conditional relationship between genes, estimating the multivariate dependence structure of expression within Tfh cells is necessary. Figure 1 illustrates the issue of zero-inflation. The data are clearly poorly modeled by the linear regression models whose fit is shown in the lower panels of the figure.

1.2. *Modeling zero-inflation.* In order to accommodate the distributional features observed in single cell gene expression, we propose a joint probability density function $f(\mathbf{y})$ of the form

$$\log f(\mathbf{y}) = \mathbf{v_y}^T\mathbf{G}\mathbf{v_y} + \mathbf{v_y}^T\mathbf{H}\mathbf{y} - \frac{1}{2}\mathbf{y}^T\mathbf{K}\mathbf{y} - C(\mathbf{G},\mathbf{H},\mathbf{K}), \quad \mathbf{y} \in \mathbb{R}^m, \tag{1}$$

for the dominating measure obtained by adding a Dirac mass at zero to the Lebesgue measure. The vector $\mathbf{y} \in \mathbb{R}^m$ comprises the expression levels of $m$ genes in a single cell, and the vector $\mathbf{v_y} \in \{0,1\}^m$ is defined through elementwise indicators of non-zero expression, so $[\mathbf{v_y}]_i = I_{\{y_i \neq 0\}}$ for $i = 1, \dots, m$. In the specification from (1), both binary and continuous versions of gene expression are sufficient statistics, and interactions thereof are parametrized, with $\mathbf{G}$, $\mathbf{H}$ and $\mathbf{K}$ being matrices of interaction parameters. Zeros in these interaction matrices indicate conditional independences (and, thus, absence of edges in a graph for a graphical model). Specifically, the $i$th and $j$th coordinate are conditionally independent if and only if all interaction matrices have their $(i,j)$ and $(j,i)$ entries zero [Lauritzen, 1996, Theorem 3.9].

As we discuss in more detail in Section 3, the model given by (1), which we refer to as the *Hurdle model*, can be shown to be equivalent to a finite mixture model of singular Gaussian distributions. In light of the observed symmetry in the positive single cell expression levels, linking the modeling of zero-inflation with Gaussian parameters for nonzero observations is both natural and convenient. This said, it is an interesting topic for future work to develop more refined models of the continuous expression arising when genes are 'on'.

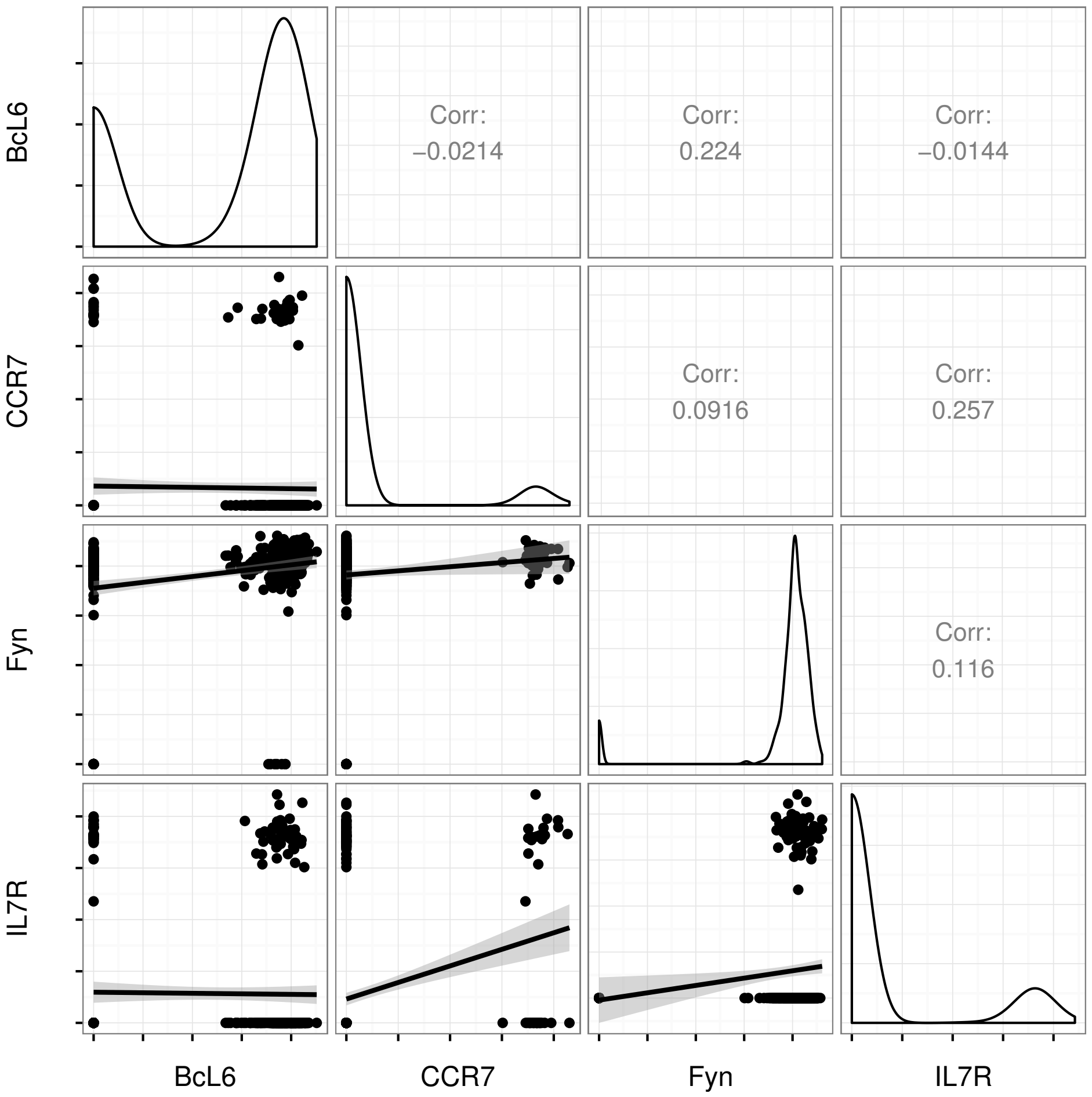

Fig 1: Scatter plots of inverse cycle threshold (40-Ct) measurements from a quantitative PCR (qPCR)-based single cell gene expression experiment (lower panels). The cycle threshold (Ct) is the PCR cycle at which a predefined fluorescence threshold is crossed, so a larger inverse cycle threshold corresponds to greater log-expression [McDavid et al., 2013]. Measurements that failed to cross the threshold after 40 cycles are coded as 0. Marginal expression in Tfh ($CXCR5^+PD1^+$) cells of Tfh marker genes is illustrated in the kernel-density estimates along the diagonal. The lower panels show the linear relationships between pairs of genes.

We will base statistical inference in the Hurdle model on so-called neighborhood selection, where the neighborhood of each gene is inferred via penalized regression methods [Meinshausen and Bühlmann, 2006]. Neighborhood selection is a state-of-the-art method for estimation and inference in potentially high-dimensional graphical models; see the review in Section 3.4 of Drton and Maathuis [2017]. The main challenge in our setting is determining how to calibrate signal in the binary versus the continuous part. We solve this problem using an *anisometric* group-lasso penalty (Section 4).

1.3. *Outline.* The remainder of the paper is structured as follows. Section 2 discusses the parameter targeted in single cell gene expression experiments, and why it is not accessible from traditional bulk experiments. Section 3 develops the parametric Hurdle model for single cell gene expression, as specified in (1), and discusses conditional independence in this setting. Section 4 gives a detailed account of estimation of graphical models using neighborhood selection via penalized regression. Section 5 provides a simulation study that demonstrates the benefits of our approach. In Section 6, we analyze the aforementioned experiment on Tfh cells. Since the data set contains selected gene profiles that were available for both single- and several-cell aggregates, we are able to highlight the refined inferences that can be obtained from single cell data. In Section 7, we analyze data on mouse dendritic cells, which are of far higher dimensionality than the Tfh cell data. Our analyses show in particular that modeling the zero-inflation may uncover distinct networks compared to existing approaches. We conclude with a discussion in Section 8, where we highlight interesting problems for future research, in particular, in graphical modeling. Section 9 contains an appendix with supplementary material and expanded derivations.

**2. Single cell versus bulk expression experiments.** Protocols for bulk gene expression experiments, such as for Illumina TrueSeq, call for 100 nanograms of total mRNA, hence require hundreds to thousands of cells. On the one hand, this biological "summation" over many of cells is expected to yield sharper inference on the mean expression level of each gene. However, it can also be expected to distort any conditional (in-)dependences present between genes.

Let $\mathbf{Y}_1, \dots, \mathbf{Y}_n$ be *iid* random vectors taking values in $\mathbb{R}^m$, with $\mathbf{Y}_i$ representing the copy numbers of $m$ transcripts present in the $i$th single cell. Now suppose the $n$ cells are aggregated, and the total expression is measured using a linear quantification that reports values proportional to the input counts of mRNA. The expression observed in this *bulk* experiment is then

$$\mathbf{Z} \propto \sum_{i}^{n} \mathbf{Y}_i,$$

with the constant of proportionality typically a semi-empirical normalization factor, such as TPM (Transcripts Per Kilobase Million) or FPKM (Fragments Per Kilobase Million). Although most bulk experiments are designed to test for differences in mean expression due to experimental treatments and lack extensive replication within a condition, *stochastic profiling* [Janes et al., 2010] experiments have provided *iid* replicates of $\mathbf{Z}$ suitable for estimating higher order moments. However, when the distribution of $\mathbf{Y}_i$ obeys some conditional independence relationships, in general the distribution of $\mathbf{Z}$ does not obey these same relationships.

For example, take $m = 3$ and suppose that the $\mathbf{Y}_i$ are *iid* samples from a tri-variate distribution supported on $\{0,1\}^3$. Let $[Y_1, Y_2, Y_3]$ be a random vector following this distribution, and let $p_{ijk} = P(Y_1 = i,\, Y_2 = j,\, Y_3 = k)$ be the joint probabilities. Then $Y_1$ and $Y_3$ are conditionally independent given $Y_2$ (in symbols, $Y_1 \perp\!\!\!\perp Y_3 \,|\, Y_2$) if and only if the two matrices $(p_{i0k})_{ik}$ and $(p_{i1k})_{ik}$ have rank 1 [Drton et al., 2009, Prop. 3.1.4]. Yet even summing over only $n = 2$ cells, the random vector $\mathbf{Z} = \mathbf{Y}_1 + \mathbf{Y}_2 \equiv [Z_1, Z_2, Z_3]$ taking values in $\{0,1,2\}^3$ generally does not have $Z_1 \perp\!\!\!\perp Z_3 \,|\, Z_2$.

When the $\mathbf{Y}_i$ are multivariate Normal, the conditional independence structure is preserved under convolution. Unfortunately for non-Gaussian distributions this does not generally hold. As noted in our introduction, single cell gene expression is generally bimodal and zero-inflated, so not plausibly described by a multivariate Normal distribution. Therefore, even though for large enough $n$ the distribution of the bulk experiment $\mathbf{Z}$ might approach multivariate (log-)normality, the networks estimated from graphical modeling of bulk data will not reflect conditional independences that hold among expression levels in single cells.

**3. Hurdle models.** Univariate Hurdle models arise from modification of a density through excision of points in the support and assignment of positive masses to these points. Targeting zero-inflation, our excision point is the origin. Let $v_y = I_{\{y \neq 0\}}$ be the indicator function for a non-zero value of the observation $y$. Then the Hurdle model derived from a Normal distribution with mean $\xi$ and precision $\tau^2$ has density

$$(2)\quad f(y) = \exp\left\{v_y\left[1/2\log\left(\tau^2/\left(2\pi\right)\right) + \log p/(1-p) - \xi^2\tau^2/2\right] + y\xi\tau^2 - y^2\tau^2/2 + \log(1-p)\right\}$$

with respect to the measure $\lambda_0$ that is the sum of the Lebesgue measure and a Dirac mass at zero. Here, $P(V_y = 1) = p \in (0,1)$ is a mixing weight representing the chance of observing a non-zero value. Varying $p$, $\xi$ and $\tau^2$, one obtains an exponential family with sufficient statistic $y$, $-y^2/2$, and $v_y$, and associated natural parameters $h = \xi\tau^2$, $k = \tau^2$, and

$$g = 1/2\log\left(\tau^2/\left(2\pi\right)\right) + \log p/(1-p) - \xi^2\tau^2/2.$$

3.1. *Multivariate Hurdle models.* A plausible model for the joint distribution of a random vector $\mathbf{Y} = [Y_1, \ldots, Y_m]$ representing single cell gene expression puts positive mass on every one of the $2^m$ coordinate subspaces (recall Figure 1), including the origin when all genes are 'off' and the entire space $\mathbb{R}^m$ when all genes are 'on'. Assigning positive mass to the coordinate subspaces generalizes the univariate construction from (2). As it is easiest to construct this model conditionally, we introduce the vector $\mathbf{V} = [V_1, \ldots, V_m]^T \equiv [I_{\{y_1\neq 0\}}, \ldots, I_{\{y_m\neq 0\}}]^T$ that indicates the non-zero coordinates of $\mathbf{Y}$. Throughout, our notation suppresses the dependence of $\mathbf{V}$ on $\mathbf{Y}$. We emphasize that specification of the distribution of the multivariate Bernoulli random vector $\mathbf{V}$ simply amounts to specification of a $2^m$ probability table.

For any vector $\mathbf{v} = [v_1, \ldots, v_m] \in \{0,1\}^m$, define the subspace $\mathbb{R}^{\mathbf{v}} = \prod_{i=1}^m \mathbb{R}^{v_i}$ where we set $\mathbb{R}^0 = \{0\}$. So, $\mathbb{R}^{\mathbf{v}}$ is the coordinate subspace corresponding to the non-zero entries of $\mathbf{v}$. Similarly, define $PD(\mathbf{v})$ to be the cone of $m \times m$ symmetric matrices that have non-zero entries only in rows and columns indexed by $i$ with $v_i = 1$, and for which the submatrix given by these rows and columns is positive definite. Now suppose that the conditional distribution of $\mathbf{Y}$ given $\mathbf{V}$ is multivariate Normal and, specifically,

$$(3)\qquad (\mathbf{Y}|\mathbf{V} = \mathbf{v}) \;\sim\; \mathcal{N}\left(\boldsymbol{\mu}(\mathbf{v}), \boldsymbol{\Sigma}(\mathbf{v})\right)$$

with mean vector $\boldsymbol{\mu}(\mathbf{v}) \in \mathbb{R}^{\mathbf{v}}$ and covariance matrix $\boldsymbol{\Sigma}(\mathbf{v}) \in PD(\mathbf{v})$. The normal distribution in (3) is singular (see Section 9.2 for details) and supported on the subspace $\mathbb{R}^{\mathbf{v}}$.

In the applications we have in mind the dimension $m$ will be large enough so that it is infeasible to accurately estimate a general $2^m$ probability table for the distribution of $\mathbf{V}$, and a collection of $2^m$ mean vectors and covariance matrices for the conditional distribution of $\mathbf{Y}$. We thus proceed to formulate a more parsimonious pairwise interaction model. While of far lower dimension, the pairwise model allows one to capture interesting conditional (in-)dependences.

First, we assume $\mathbf{V}$ to follow an Ising model with joint probabilities

$$p(\mathbf{v}) \equiv P(\mathbf{V} = \mathbf{v}) \propto \exp\left(\mathbf{v}^T \mathbf{G} \mathbf{v}\right), \qquad \mathbf{v} \in \{0,1\}^m, \tag{4}$$

where $\mathbf{G}$ is a symmetric interaction matrix in $\mathbb{R}^{m \times m}$. Second, we assume that the conditional normal distribution of $\mathbf{Y}$ given $\mathbf{V} = \mathbf{v}$ has log-density

$$\log f(\mathbf{y}|\mathbf{V} = \mathbf{v}) = \mathbf{v}^T \mathbf{H} \mathbf{y} - \frac{1}{2}\mathbf{y}^T \mathbf{K} \mathbf{y} - C'(\mathbf{H}, \mathbf{K}), \qquad \mathbf{y} \in \mathbb{R}^{\mathbf{v}}, \tag{5}$$

with respect to Lebesgue measure restricted to the subspace $\mathbb{R}^{\mathbf{v}}$. In (5), $\mathbf{H}$ and $\mathbf{K}$ are two $m \times m$ interaction matrices that do not vary with $\mathbf{v}$, and $C'(\mathbf{H}, \mathbf{K})$ is a normalization constant. The matrix $\mathbf{K}$ is symmetric and positive definite, but $\mathbf{H}$ may be arbitrary from $\mathbb{R}^{m \times m}$. Putting the two pieces from (4) and (5) together, the joint density of $\mathbf{Y}$ with respect to the product measure $\lambda_0^m$ simplifies to

$$f(\mathbf{y}) = \exp\left\{\mathbf{v}^T \mathbf{G} \mathbf{v} + \mathbf{v}^T \mathbf{H} \mathbf{y} - \frac{1}{2}\mathbf{y}^T \mathbf{K} \mathbf{y} - C(\mathbf{G}, \mathbf{H}, \mathbf{K})\right\}, \qquad \mathbf{y} \in \mathbb{R}^m. \tag{6}$$

We recognize an exponential family with three interaction matrices $\mathbf{G}$, $\mathbf{H}$ and $\mathbf{K}$ as natural parameters and the three statistics $\mathbf{v}\mathbf{v}^T$, $\mathbf{v}\mathbf{y}^T$, and $\mathbf{y}\mathbf{y}^T$ sufficient.

Let $\mathcal{I} \equiv \mathcal{I}(\mathbf{V})$ be the $m \times m$ diagonal matrix with $(i,i)$ entry equal to $V_i$. Then for any vector $\mathbf{x} \in \mathbb{R}^m$ the product $\mathcal{I}\mathbf{x}$ is the vector that has the $i$th coordinate replaced by zero for all indices $i$ with $Y_i = V_i = 0$. Similarly, multiplying $\mathcal{I}$ from left and right to a matrix zeros out all but the principal submatrix determined by this set of indices. Using this notation, the pairwise Hurdle model from (6) corresponds to the particular choice of

$$\boldsymbol{\mu}(\mathbf{v}) = \left(\mathcal{I}\mathbf{K}\mathcal{I}\right)^{-} \mathbf{H}\mathbf{v}, \qquad \boldsymbol{\Sigma}(\mathbf{v}) = \left(\mathcal{I}\mathbf{K}\mathcal{I}\right)^{-} \tag{7}$$

for the mean vectors and covariance matrices in the conditional specification from (5). In (7), $A^-$ denotes the Moore-Penrose pseudoinverse of a matrix $A$. From the perspective of (7), the pairwise Hurdle model is a mixture of $2^m$ singular Gaussian distributions whose mean vectors and covariance matrices are derived from one precision matrix $\mathbf{K}$ and an interaction matrix $\mathbf{H}$.

The notation we used in the conditional specification of the multivariate Hurdle model follows Lauritzen [1996], who describes conditional Gaussian (CG) models with *inhomogeneous, non-singular* precision $\mathbf{K}(\mathbf{v})$ that can depend on the discrete set of covariates in arbitrary, positive-definite fashion. These models have been considered more recently by Lee and Hastie [2013] and Cheng et al. [2013]. Our formulation differs from the traditional CG models by involving singular distributions with means and covariance matrices that exhibit structured inhomogeneity.

3.2. *Conditional distributions identify interaction parameters.* The normalizing constant $C$ in equation (6) is a difficult to compute sum of $2^m$ terms. This is expected as already the distributions in the Ising model from (4) have an intractable normalization constant for moderately large $m$. Fortunately, the univariate full conditional distributions obtained from (6) have tractable normalizing constants and identify the parameters from a given row/column of the interaction matrices $\mathbf{G} = (g_{ab})$, $\mathbf{H} = (h_{ab})$, and $\mathbf{K} = (k_{ab})$.

Fix a coordinate $b$, and define its complement $A = \{1, \ldots, m\} \setminus \{b\}$. Consider now the density $f(\mathbf{y})$ from (6) as a function of only $y_b$, i.e., $\mathbf{y}_A = [y_i : i \in A]$ is fixed, and write $f_{[b|A]}$ for the conditional density of $y_b$ given $\mathbf{y}_A$. Then noting that $v_i y_i = y_i$ and $v_i^2 = v_i$, we have

$$\log f_{[b|A]}(\mathbf{y}) = v_b g_{[b|A]} + y_b h_{[b|A]} - \frac{1}{2} y_b^2 k_{[b|A]} - C_{[b|A]}, \qquad y_b \in \mathbb{R}, \tag{8}$$

where $C_{[b|A]}$ does not depend on $y_b$ and

$$g_{[b|A]} = g_{bb} + 2\mathbf{g}_{bA}\mathbf{v}_A + \mathbf{h}_{bA}\mathbf{y}_A, \tag{9}$$

$$h_{[b|A]} = h_{bb} + \mathbf{h}_{Ab}^T\mathbf{v}_A - \mathbf{k}_{bA}\mathbf{y}_A, \tag{10}$$

$$k_{[b|A]} = k_{bb}.$$

The conditional density $f_{[b|A]}$ is thus a univariate Hurdle density as specified in (2) with natural parameters $g_{[b|A]}$, $h_{[b|A]}$, and $k_{[b|A]}$.

The three natural parameters are obtained from linear predictors that depend on a design matrix constructed from $\mathbf{y}_A$ and $\mathbf{v}_A$. For example, we may write

$$g_{[b|A]} = g_{bb} + \sum_{a \in A} X_a \begin{bmatrix} g_{ba} \\ h_{ba} \end{bmatrix}$$

for $X_a = [v_a, y_a]$. The linear predictor for $h_{[b|A]}$ can be written analogously. We note that if the data include additional nuisance covariates $\mathbf{W}_0$ that describe each experimental unit then these can be included by augmenting the linear predictor to

$$g_{[b|A]} = \mathbf{W}_0^T \mathbf{g}_{b0} + g_{bb} + \sum_{a \in A} X_a \begin{bmatrix} g_{ba} \\ h_{ba} \end{bmatrix} \tag{11}$$

with $\mathbf{g}_{b0}$ being the parameters capturing the effects of the covariates. From this perspective, the conditional distribution in (8) defines a vector generalized linear model, parametrized by three natural parameters $g_{[b|A]}$, $h_{[b|A]}$ and $k_{[b|A]}$, the first two of which are modeled as a linear function of the expression of other genes.

3.3. *Conditional independence graphs.* The dependence structure of the random vector $\mathbf{Y} = [Y_1, \ldots, Y_m]$ may be summarized in its *conditional independence graph.* This is an undirected graph $\mathcal{G} = (\mathcal{V}, \mathcal{E})$ with vertex set $\mathcal{V} = \{1, \ldots, m\}$ and an edge set $\mathcal{E}$ that is determined by the conditional independences in $\mathbf{Y}$. More precisely, the edges in $\mathcal{E}$ are those two-element sets $\{a, b\} \subset \mathcal{V}$ for which $Y_a$ and $Y_b$ are conditionally dependent given the remaining variables, i.e., $\mathbf{Y}_{\mathcal{V}\setminus\{a,b\}}$. In our case, $\mathbf{Y}$ has a density $f$ as in (6). The dominating measure is a product measure, and $f$ is positive and continuous. Hence, the Hammersley-Clifford theorem assures that the conditional independence graph of $\mathbf{Y}$ has an edge $\{a, b\}$ if and only if the four possible $ab$ interactions are zero, so

$$g_{ab} = h_{ab} = h_{ba} = k_{ab} = 0; \tag{12}$$

see Lauritzen [1996, Chapter 3]. This fact is also evident from the form of the conditional distributions detailed in (8), (9), and (10). It motivates the neighborhood selection procedure developed in the next section.

**4. Neighborhood estimation via penalized regression.** In the single cell experiments to which we envision applying this method, the number of cell replicates, $n$, is larger than the sample sizes seen in typical bulk mRNA experiments. However, it is still often the case that the number of genes $m$ is larger than the number of cell replicates. We are thus in a setting that benefits from application of methods from 'high-dimensional statistics'; though emerging technologies are increasing available sample sizes.

4.1. *Related work.* Under scenarios in which $n, m \to \infty$ while satisfying that $n > Cd^{\phi}(\log m)^{\psi}$, where $C, \phi$ and $\psi$ are constants that depend on the model and $d$ is the maximum vertex degree of the conditional independence graph, penalized regression has been shown to consistently identify the graph of multivariate Normal models [Meinshausen and Bühlmann, 2006], of Ising (auto-logistic) models [Ravikumar et al., 2010] and of exponential family graphical models [Yang et al., 2014, Chen et al., 2015]. While this paper was in preparation, Tansey et al. [2015] further extended this line of work to general *vector space graphical models* that include the multivariate Hurdle model as a special case. However, the standard (isometric) group-lasso they propose for estimation of the conditional independence graph does not account for heterogeneity in the scaling of predictors in the conditional distributions. The anisometric group-lasso we propose in the following section yields drastic improvements in finite samples.

4.2. *Anisometric penalty.* Throughout this section, we fix an index $b$ and consider the conditional distribution $Y_b$ given the other variables in $\mathbf{Y}_A$ for $A = \{1, \ldots, m\} \setminus \{b\}$. For any $a \in A$, define the parameter vector $\theta_a = [g_{ba}, h_{ba}, h_{ab}, k_{ba}]$. By (12), $Y_b \perp\!\!\!\perp Y_a | \mathbf{Y}_{A\setminus\{a\}}$ if and only if $\theta_a = 0$.

Let $\theta = [\theta_a : a \in A]$, and let

$$P_\lambda(\theta) = \lambda \sum_{a \in A} \sqrt{\theta_a^T \theta_a} \tag{13}$$

be the group lasso penalty for tuning parameter $\lambda \geq 0$. Maximization of the penalized conditional log-likelihood function

$$\log f_{[b|A]}(\mathbf{y}) - P_\lambda(\theta)$$

can lead to a solution that is sparse in parameter blocks, that is, some of the subvectors $\theta_a$ are zero. The penalty is equivalent to placing a sequence of independent, multivariate Laplace priors on blocks of $\theta$ and reporting the MAP [Eltoft et al., 2006].

Viewed as a prior, the standard group-lasso penalty from (13) implicitly assumes that each variable in each block has a similar effect size. This may be reasonable if the variables in each block are measured in comparable units, but is problematic otherwise. For example, if covariate $X_1$ is measured in meters, while covariate $X_2$ in centimeters, then the distribution of effect sizes for $X_2$ would be 100-times more dispersed than the distribution of effect sizes for $X_1$. In penalized GLMs, this is typically enforced "at run time" by ensuring covariates are on comparable scales, or Z-scoring each column of the design matrix if no intrinsic scale exists.

In our setting of a vector regression, terms from linear predictor $g_{[b|A]}$ and linear predictor $h_{[b|A]}$ end up together in blocks, and these coefficients are not necessarily comparable, as one specifies log-odds of $E(V_b|V_A = 0)$ while the other specifies conditional expectations of $E(Y_b|Y_A)$. Re-scaling does not resolve this, since the same design matrix $X_a = [V_a, Y_a]$ is used in each linear predictor, and in any case, re-scaling generally alters the solution [Simon and Tibshirani, 2012]. Instead, we propose replacing the isometric $\ell_2$ norm in the sum in (13) so that the penalty is

$$P_{\mathbf{H},\lambda}(\theta) = \lambda \sum_{a \in A} \sqrt{\theta_a^T \mathbf{H}_{aa} \theta_a}. \tag{14}$$

Here, $\mathbf{H} \equiv \text{diag}(\mathbf{H}_{aa})$ is a block-diagonal, positive-definite matrix that allows terms from the linear predictors to have different scales of penalty. It also accounts for correlation between components of $\theta_a$, since columns of the design are correlated due to both $v_a$ and $y_a$ appearing as predictors.

If prior information existed, the matrix $\mathbf{H}$ could be chosen accordingly, with interpretation as a multivariate Laplace prior. Absent prior information, setting $\mathbf{H}$ equal to the Fisher information under a null model $\theta_a = 0$ for all $a$ results in variable selection approximately equal to conducting score tests, with exact equivalence holding under a null hypothesis of $\theta_a = 0$ for all $a$; see Proposition 1 in Section 9.

4.3. *Computation.* In Algorithm 1, we outline the proposed neighborhood selection, allowing for possible nuisance covariates $\mathbf{W}$. The nuisance covariates $\mathbf{W}$ might just be an intercept column, but generally could be any cell-level covariate deemed relevant. The smooth and concave function in line 7 can be maximized using any Newton-like algorithm (e.g., BFGS). The objective in line 10 is a sum of a concave, smooth function and a structured concave function and can be efficiently solved using proximal gradient ascent [Parikh and Boyd, 2014]. In particular, one may exploit the fact that although the proximal operator

$$\text{prox}_\gamma(x) = \text{argmax}_u \frac{1}{\gamma}\|x-u\|_2^2 + \sum_{a\in A}\sqrt{u_a^T H_{aa} u_a}$$

is not available in the familiar form of a soft-thresholding operator as in the isometric group-lasso, the proximal operator of the *anisometric* group-lasso can be efficiently found via a line search after one-time pre-calculation of the singular value decomposition of $H_{aa}$ [Foygel and Drton, 2010]. Throughout the inner-loop, warm starts are exploited for $\hat{\theta}$ as $\lambda$ varies. Active set heuristics using the strong rules of Tibshirani et al. [2012] yield computational gains for sparse solutions with large $m$. The algorithm yields, for each node, a sequence of neighborhoods over a sequence of tuning parameters $\mathbf{\Lambda}$. These neighborhoods need not be consistent, in the sense that for some element of $\mathbf{\Lambda}$ it could be that $b \in \text{Ne}(a)$ but $a \notin \text{Ne}(b)$. We resolve that by adopting an "or" rule. In the accompanying software[1], the algorithm is written in a combination of `R` and `C++`. Timings for the proposed method and competitors (described further in Section 5) are shown in Figure 3.

**5. Simulations.** We consider a series of simulations under several sets of underlying i) graph topologies, ii) parametric models, iii) sample sizes and

[1]Available `https://github.com/amcdavid/HurdleNormal`

**Data:** Expression matrix $\mathbf{Y} \in \mathbb{R}^{n \times m}$, nuisance covariates $\mathbf{W} \in \mathbb{R}^{n \times q}$, penalty path $\mathbf{\Lambda}$.
**Working parameters:** Unpenalized nuisance parameters $\theta_0 \in \mathbb{R}^{2q+1}$, edge parameters $\theta \in \mathbb{R}^{4(m-1)}$.
**Result:** Neighborhoods $ne(i, \lambda)$, $1 \leq i \leq m$, $\lambda \in \mathbf{\Lambda}$

**for** $b \in \{1, \ldots, m\}$ **do**
  $A \leftarrow \{1, \ldots, m\} \setminus \{b\}$ ;
  $\mathbf{X} \leftarrow \left[\mathbf{W}, \mathbf{Y}_A, \mathbf{V}_A\right]$ ;
  $\theta_0 \leftarrow \left[g_{bb}, \mathbf{g}_{b0}^T, h_{bb}, \mathbf{h}_{b0}^T, k_{bb}\right]$ ;
  $\theta \leftarrow \left[\mathbf{g}_{bA}, \mathbf{h}_{bA}, \mathbf{h}_{Ab}^T, \mathbf{k}_{bA}\right]$ ;
  Let $\log f_{[b|A]}\left(\theta_0, \theta\right)$ return the log-density (8) evaluated at $[\theta_0, \theta]$ with covariate matrix $\mathbf{X}$.
  $\bar{\theta}_0 \leftarrow \text{argmax}_{\theta_0} \log f_{[b|A]}(\theta_0, \theta = 0)$ ;
  $\mathbf{H} \leftarrow \nabla^2 \log f_{[b|A]}\left(\bar{\theta}_0, 0\right)$;
  **for** $\lambda \in \mathbf{\Lambda}$ **do**
    $[\hat{\theta}_0, \hat{\theta}] \leftarrow \text{argmax}_{\theta_0, \theta} \log f_{[b|A]}(\theta_0, \theta) - P_{\mathbf{H},\lambda}(\theta)$ ;
    Let $ne(b, \lambda)$ contain vertex $a$ whenever any of $\hat{\mathbf{g}}_{bA}, \hat{\mathbf{h}}_{Ab}, \hat{\mathbf{h}}_{bA}, \hat{\mathbf{k}}_{Ab} \neq 0$.
  **end**
**end**

**Algorithm 1:** Neighborhood selection

iv) number of vertices. We summarize the considered setups here and defer details (including the choice of graph topology) to the supplementary material in Section 9.5. The number of observations $n$ varies from 100 to 12500. In the *chain* graph topology, the number of vertices varies from $m = 16$ to $m = 128$, while in the *e. coli* graph topology, $m = 500$. The parametric models include the pairwise hurdle model (6), the hurdle model under contamination by $t_8$ noise, a logistic/Ising model and a Gaussian/logistic censoring model specified in Equation (15) in Table 1. The pairwise hurdle model is said to be *complete* if for each edge present in the graph, all of the corresponding entries in each of the three interaction matrices are non-zero. The pairwise hurdle model is said to be *G-minimal* when $H$ and $K$ are diagonal matrices and only $G$ contains non-zero off-diagonal entries. In this case, the G-minimal model is equivalent to a logistic/Ising model.

5.1. *Methods compared and default tunings.* Six methods were examined to test graph structure inference, and are described in Table 1. The Hurdle models are fit using the accompanying software `HurdleNormal` version 0.98.2, while the Logistic, Gaussian and NPN models are fit using the `R` package `glmnet` version 2.0-5 (via the `autoGLM` function in `HurdleNormal`). The Aracne method is fit using package `netbenchmark` version 1.6.0. For methods 1-5, neighborhoods are stitched together using an "or" rule, i.e., vertices $a$ and $b$ are adjacent if either $b \in \text{ne}(a)$ or $a \in \text{ne}(b)$.

In Figure 4 various fixed tunings are shown. In the *oracle* tuning, the graph with maximum sensitivity subject to FDR $< 10\%$ is shown. This tuning is not available in practice, but shows the maximum achievable performance of each method. With the *BIC* tuning, we employ the Bayesian Information Criteria on the pseudo-likelihood

$$\text{BIC}_\lambda = \sum_{b \in \mathcal{V}} -2 \log f_{[b|A]}(\hat{\theta}_{b,\lambda}) + \|\theta_{b,\lambda}\|_0 \log n,$$

where $\theta_{b,\lambda}$ is the penalized solution at penalty $\lambda$ for vertex $b$, $\|\theta_{b,\lambda}\|_0$ is the number of non-zero entries, and $\hat{\theta}_{b,\lambda}$ is the (unpenalized) maximum pseudo-likelihood estimate for the non-zero entries. The BIC solution is the one that minimizes $\text{BIC}_\lambda$. This tuning is available for methods 1-5. In the case of the the Aracne method the BIC is unavailable as no likelihood is defined.

5.2. *Results.* 30 simulation replicates sufficed to bound the simulation-induced Monte Carlo standard error of the mean $< 5 \times 10^{-3}$ for FDR and $< .02$ for the sensitivity.

The simulations show that mis-specified estimation procedures perform poorly when model (6) is the data generating distribution. When an FDR-controlling oracle is available, the anisometric Hurdle model can dominate other methods in edge-sensitivity (Figure 4A-B). However, when the Hurdle model is over-parameterized as in the G-sparse scenarios, the minimal Logistic model is superior, though the anisometric $\ell_1$ penalty partially ameliorates this gap. In very simple chain-graph scenarios, it is neigh-impossible to recover a network using 10-cell data. The *e. coli* network provides a counter example where 10-cell data nearly equals the performance available from single cell data. This may be due to the hub-and-spoke nature of the *e. coli* network, so the effect of *marginalization by convolution* tends to only add more connections between the hub and its neighborhood. The *e. coli* data and chain-graphs suggest that collecting single cell data, and estimating graph structure with a method that accommodates zero inflation can accurately discover a wide variety of network topologies.

More seriously, ignoring zero-inflation confounds use of information criteria to tune network size (panel C). On the other hand, the Hurdle model is robust to a variety of model departures, including contamination with $t_8$-distributed errors (labelled with "t"), and data generation under a Gaussian-Logistic censoring model. When the full solution path is examined (Figure 5), a practioner who reported only the top few edges would often suffer from a large number of false positives when using methods not designed for

**Graph topologies and parametric models**

1. $G$-minimal chain graphs, with tri-diagonal $G$-interaction matrix with off-diagonal entries set to 1, and diagonal H, K. In this case, an Ising/logistic model is minimally complete.
2. $G$-$H$-$K$-complete chain graphs, with off diagonal $G = .2, H = -.75, K = -.4$. The proposed model is thus minimally complete.
3. *e. coli*-networks: 500 edges from a semi-empirical *e. coli* network and pairwise hurdle likelihood. 50% of edge weights are $G$-minimal, 25% $K$-minimal and 25% complete.
4. 10-cell versions of 1-3. The 10-cell observation $\mathbf{Y}^{(\mathbf{10})}$ is generated as $\mathbf{Y}^{(\mathbf{10})} = \log_2 \sum_{i=1}^{10} 2^{\mathbf{Y}_i}/10$ and $\mathbf{Y}$ is generated as under model 1-3.
5. 1-3 with non-zero observations contaminated with $t_8$ noise.
6. 1-3 with the following latent Gaussian/logistic selection model:

$$\tilde{\mathbf{Y}} \sim \mathcal{N}(\mu, \mathbf{K}),$$
$$P\left(\tilde{V}_j | \tilde{\mathbf{Y}} = \tilde{\mathbf{y}}\right) = \text{logit}(a + b\tilde{y}_j),$$
$$\mathbf{Y} = \tilde{\mathbf{Y}}\tilde{V}. \tag{15}$$

**Methods**

1. Aracne [Margolin et al., 2006]: connects genes with significant pairwise mutual information and applies pruning rules to suppress indirect effects.
2. Gaussian: neighborhood selection with $\ell_1$-penalized linear regression [Meinshausen and Bühlmann, 2006].
3. Logistic: neighborhood selection with $\ell_1$-penalized logistic regression[Ravikumar et al., 2010].
4. NPN: neighborhood selection with $\ell_1$-penalized linear regression on Gaussian-quantile transformed responses [Liu et al., 2009].
5. Hurdle (isometric): neighborhood selection with model (6) and isometric group-lasso penalty.
6. Hurdle (anisometric): neighborhood selection with model (6) and anisometric group-lasso penalty.

Table 1
*Overview of simulation scenarios and methods compared.*

zero-inflated data. For example, with $n = 100$ in the *e. coli* network, all methods, aside from the Hurdle have FDR exceeding 20%. The simulations also suggest that perfect recovery of gene networks is impractical at realistic sample sizes, even with a correctly specified model, motivating a form of meta-analysis on estimated graphs, discussed further in Section 7.2.

**6. T follicular helper cells.** Our simulations show that depending on the data generating scenario, the Hurdle method may substantially outperform, or at least mimic the performance of other candidate methods. We next sought to see if methods would tend towards consensus in biologically-derived single cell and 10-cell data, or if it were possible that the Hurdle method might offer unique insights. We considered co-expression networks in Tfh cells measured in eight healthy donors. 65 genes were selected for profiling via qPCR on the basis of their role in Tfh signaling and differentiation, generally with sparse expression across single cells (overall probability of expression 27%). 465 single cell, and 187 10-cell replicates were taken.

Figure 6 shows networks of approximately 24 edges estimated using Hurdle, Gaussian (with centered data, see Section 9) and Logistic, and Gaussian model using 10-cell aggregates. The size of the network is a compromise between stability selected [Shah and Samworth, 2013] sizes of each procedure, which varies from 11 edges (Hurdle) to 32 edges (Gaussian).

Normalized Hamming distances between the four methods, the Aracne method and the Gaussian model fit on the "raw", uncentered data are reported in Table 2. The Hurdle and Gaussian models are most similar, while the logistic and Gaussian 10-cell network are quite distinct. The Gaussian(raw) model on untransformed data is similar to the logistic model, as distance of non-zero expression values from the origin is large compared to the variation among the non-zero values.

In the Hurdle network, the transcription factors NFATC1 (Nuclear factor of activated T-cells) and BCL6, and the signaling molecule CD154 and chemokine receptor CCR3 are hubs. NFATC1 has been found to promote transcription of cytokines IL21 [Hermann-Kleiter and Baier, 2010] and signaling molecule CD154 [Pham et al., 2005], while BCL6 serves as a transcriptional repressor, and is one of the canonical markers constitutively expressed in Tfh cells. CTLA4 which has been described to inhibit inflammation, interacts negatively with inflammatory activator JAK3. The disconnected component of CCR3-CCR4-BTLA-SELL-TNFSF4 may hint at plasticity between Tfh cells and the related T-cell lineages Th1 and Th2. CCR3 and CCR4 are canonical markers of Th2 cells, while TNFSF4 (coding for OX40L) promotes Th2 development de Jong et al. [2002]. Thus co-expression of these

| | Gaussian (10) | Gaussian | Gaussian(raw) | Hurdle | logistic |
|---|---|---|---|---|---|
| Aracne | 1.00 | 0.92 | 0.92 | 1.00 | 1.00 |
| Gaussian(10) | | 1.00 | 1.00 | 1.00 | 1.00 |
| Gaussian | | | 0.92 | 0.65 | 1.00 |
| Gaussian(raw) | | | | 1.00 | 0.39 |
| Hurdle | | | | | 1.00 |

TABLE 2

*Dissimilarities* $\left(\frac{\textit{Hamming Distance}}{\textit{Number of edges}}\right)$ *between networks of size 24 estimated through various methods. The Gaussian(10) model is a Gaussian model estimated on 10-cell replicates, while the Gaussian(raw) data is estimated on single cells without centering the data. The remaining models are described in Section 5.*

genes may suggest cells transitioning between Tfh and Th1 or Th2 states.

In the Gaussian network, though NFATC1, BCL6 and CD154 remain highly connected, CD27 now has highest degree and serves as a hub to receptors CXCR4, IL2Rb, IL2Rg, as well as ITGB2, NFATC1 and FYN. CD3e, the backbone responsible for transducing the T-cell receptor signal is connected with co-receptor CD4, CD154, IL2Rg, Fyn and ANP32B. The negative interactions between BTLA and CTLA4 are absent.

The logistic network consists primarily of negative interactions. The strongly negative BCL6–BLIMP1 edge is consistent with previously described antagonism between these genes [Johnston et al., 2009]. Interestingly, this edge is absent in the other networks. Networks found by applying the Aracne and Gaussian(raw) methods are shown in the supplementary material.

## 7. Mouse dendritic cells.

Shalek et al. [2014] exposed bone marrow-derived dendritic cells, from *mus musculus*, to lipopolysaccharide (LPS). LPS is a toxic compound secreted and structurally utilized by gram-negative bacteria and induces a cascade of changes in a cell's expression profile through several pathways. Cells were sampled after 0, 1, 2, 4, and 6 hours post-exposure. We estimated transcription networks using 4431 transcripts expressed in at least 20% of 65 cells sampled 2 hours after LPS exposure, at which interval transcription is expected to be undergoing a variety of dynamic changes. Rather than attempting to perform model selection on this limited sample size, we consider highly sparse ($< .01\%$ sparsity) networks of 700 edges, chosen to provide tractable visualization and illustration of the method. The BIC tunings (discussed subsequently) are decidedly larger.

### 7.1. *Selected networks.*

In a Gaussian model, the network is star-shaped, with Mx1, Ccl17, Tax1bp3 and Ccl3 as hubs all with degrees $\geq 15$, though none are directly inter-connected (Figure 7). In all, 2.5% of non-isolated vertices contribute 50% of the edges in the network. With the exception of Tax1bp3, these hub genes are all immune-signaling related.

In the Hurdle model (Figure 8), the graph is more chain-like, with maximum degree 12: 7% of nodes provide 50% of the edges. The strongest hub, Mgl2 (also known as Cd301b), has been recently described to be involved in uptake and presentation of glycosylated antigens, such as LPS, by dendritic cells [Denda-Nagai et al., 2010]. A sub-connected set of genes coding for MHC-II antigen presentation (H2ab1, H2eb1, H2aa) is the densest subcomponent, and interconnected to Mgl2 as well as Fabp5. Increased expression of Fabp5 has been shown to increase expression of cytokines Il7 and Il18, hence is also involved in immune cell stimulation [Adachi et al., 2012]. Many of the neighbors of Mgl2,H2ab1, H2eb1, H2aa and Fabp5 are neighbors of the hub genes in the Gaussian graph, whereas Mx1, Ccl17 and Ccl3 are sparsely connected in the Hurdle network. Tax1bp3 is absent.

Using BIC, both the Gaussian and Logistic models yield networks with more than 25,000 edges, while the Hurdle selects a network of roughly 12,000 edges. The additional flexibility available in the Hurdle for modeling internode relationships may permit sparser graphs to describe the conditional dependence relationships. We also observe that the Hurdle synthesizes signal from both Gaussian and Logistic networks. For sufficiently rich network sizes, the Gaussian and Hurdle and Logistic and Hurdle networks share 21% and 1% of possible edges, respectively, compared to only .08% of possible edges between the Gaussian and Logistic networks (binomial test $p < 10^{-6}$).

7.2. *Graphical geneset edge enrichment.* We consider how well the 700 edge networks recapitulate known relationships between genes using previously described functional annotations. The Gene Ontology Consortium [2015] provides a database of categories to which genes may be annotated if experimentally or computationally they are involved in a biological process. We note that networks may exhibit *intraconnection* within GO categories, and that some pairs of categories may exhibit preferential *interconnection.*

Each pair $(i, j)$ of GO categories—including self-pairs—induces a *coloring* of vertices, coloring the vertices belonging to category $i$ color $c_i$ and category $j$ color $c_j$. Vertices that do not belong to either $i$ or $j$ remain uncolored. Iterating through the $3987^2/2$ pairs of categories, we test for edge enrichment between colors. Suppose in the inferred graph of 700 edges, $n_{ij}$ edges connect $c_i$-colored vertices to $c_j$ vertices. If the colored vertices were completely connected with $n_i$ vertices of color $c_i$ and $n_j$ vertices of color $c_j$, then there would be $m_{ij} = n_i \times n_j$ edges among them (with the obvious adjustment made for self-edges when $i = j$). We now define an enrichment statistic as the hypergeometric tail probability

$$t_{ij} = P(N_{ij} > n_{ij}; 700, m_{ij}, 4431 \times 4430/2),$$

which is the probability of drawing $n_{ij}$ colored balls, given 700 draws from urn containing $4431 \times 4430/2$ balls of which $m_{ij}$ are colored.

This results in nearly 16 million enrichment statistics on the pairs of categories, which follow a complicated dependence structure under the series of null hypotheses that the observed edges being connected independent of coloring. The top 200 (smallest in magnitude) enrichment statistics $t_{(k)}$, $k < 200$ are compared to their distribution $P(t^*)$ under a Erdos-Renyi random graph model, yielding a Monte Carlo p-value for each order statistic. A pair of colors $(i, j)$ with rank $r_{ij} < 200$ is declared significant if $P(t_{ij} < t^*_{(r_{ij})}) < .05$ and $P(t_{(r)} < t^*_{(r)}) < .05$ for all $r < r_{ij}$, that is, it is significant at 5% and all smaller order statistics are also significant.

7.2.1. *Hurdle graphs tend to include intra-category enrichment.* In the Gaussian model, more than 100 pairs of categories (colors) are significantly enriched at an FDR of less than 10%, however in these pairs, only 6 correspond to intra-category enrichment (Figure 10). These are: response to salt stress, potassium channel regulator activity, extracellular exosome and three genesets containing genes with significant time-course differential expression in the original experiment. In the Hurdle model (Figure 11), 13 of 57 significantly enriched pairs form intra-connections, including defense response to Gram-negative bacteria, and cell-cell adhesion and several modules involving extracellular secretion via the Golgi apparatus. Also of particular note, genes annotated to the activation of innate immune response are directly connected to RNA PolII transcription factors, as well as "detection of lipopolysaccharide"–"endoplasmic reticulumGolgi intermediate compartment." Both of the modules are absent from the Gaussian network. This suggests that the more appropriate Hurdle model manages to identify transcription factor-induced expression changes in these regulated genes, a direct method by which one gene would induce expression changes in another.

No significant enrichment was found in the logistic model.

**8. Discussion.** Graphical models estimated from single cell data are distinct from networks estimated from bulk data, or even repeated stochastic samples. In simulations, the Hurdle model with anisometric penalty has much greater sensitivity compared to available methods, while in the two data sets here, it yields substantially different network estimates compared to Gaussian and Logistic models on these zero-inflated data. When enrichment of gene ontology categories is considered between vertices in transcriptome-wide data, the enrichment uncovered with the Hurdle model is consistent with identifying direct effects of transcription factors on genes undergoing dynamic regulation due to LPS exposure.

In our work, we have utilized methods for sparse neighborhood selection. However, the zero-inflated parametric model explored here is not limited to this framework, and could serve as a basis for many network inference techniques, including mutual information-based techniques, or to parametrize families of directed networks.

Although measuring transcriptome-wide data allows conditional estimation of direct effects between genes, non-mRNA factors may also greatly affect gene expression. In this sense, important variables have still been marginalized over, and in the case of the Tfh data, indeed, most of the transcriptome has been marginalized over. Extensions that adapt graphical model selection to clustering and/or factor analytic models would likely be useful and allow greater biological insight with these data sets.

## 9. Supplementary material.

9.1. *Data processing.* The method, and code to reproduce results in this paper is available as an R package at `https://github.com/amcdavid/HurdleNormal`.

In all models and data sets, the cellular detection rate $\sum_j I_{y_{ij}>0}$ [Finak et al., 2015] was used as an unpenalized adjustment covariate in $\mathbf{W}$ as described in Algorithm 1. In the Tfh data, a separate, unpenalized intercept was fit for each donor, as well. For the Gaussian and Hurdle models, positive values were conditionally centered

$$\tilde{y}_{ij} = \begin{cases} 0 & \text{if } v_{ij} = 0, \\ y_{ij} - \bar{y}_j^+ & \text{else,} \end{cases}$$

where $\bar{y}_j^+$ is the average in a gene over positive values. This made $V_j$ and $Y_j$ marginally orthogonal, speeding up the convergence of the optimization algorithm and reducing the leverage of zeros in the Gaussian model. The "Gaussian(raw)" model was also fit to the untransformed data, but not always discussed as it gave similar results as the Logistic model.

The graph stability (via repeated 50% sample splitting) was used to estimate the network size. At 60% stability, the number of selected edges ranged from 11 (Hurdle) to 32 (Gaussian).

Background noise in the mouse dendritic cells (mDC) data set was thresholded as described previously [Finak et al., 2015], and filtered for low-expression and cluster-disrupted cells. Figure 12 shows the Bayesian information criterion for the fitted path. An interior minimum fails to occur in the solution path for three of the methods.

9.2. *Singular normal distributions.* A random vector $\mathbf{Y}$ has singular Normal distribution $\mathcal{N}(\boldsymbol{\mu}, \boldsymbol{\Sigma})$ [Rao, 1973] with mean $\boldsymbol{\mu}$ and covariance $\boldsymbol{\Sigma}$ with rank $r < m$ if the following holds for a matrix $\mathbf{U}$ with $\mathbf{U}^T\boldsymbol{\Sigma} = 0$: a) $\mathbf{U}^T\mathbf{Y} = \mathbf{U}^T\boldsymbol{\mu}$ almost surely, and b) $\mathbf{Y}$ has a density

$$f(\mathbf{y}) = \frac{(2\pi)^{-r/2}}{(\det^+ \boldsymbol{\Sigma})^{1/2}} \exp\{-(\mathbf{y} - \boldsymbol{\mu})^T \boldsymbol{\Sigma}^- (\mathbf{y} - \boldsymbol{\mu})/2\}, \tag{16}$$

with respect to Lebesgue measure restricted to the hyperplane $\mathbf{U}^T\mathbf{Y} = \mathbf{U}^T\boldsymbol{\mu}$. Here $\det^+$ is the pseudo-determinant (product of non-zero eigenvalues) and $\boldsymbol{\Sigma}^-$ is a pseudo-inverse, such as the Moore-Penrose inverse. In the case that $\boldsymbol{\Sigma}$ is zero outside a positive-definite submatrix of size $r \times r$, $\mathbf{U}$ can be chosen to be a diagonal selection matrix consisting of zeros and ones, and $\mathbf{Y}$ has a density with respect to the measure $\lambda^r \otimes \delta_0^{m-r}$, which is the case treated here.

9.3. *Normalizing the joint density.* The expression

$$f(\mathbf{y}) \;=\; \exp\left\{\mathbf{v}^T\mathbf{G}\mathbf{v} + \mathbf{v}^T\mathbf{H}\mathbf{y} - \frac{1}{2}\mathbf{y}^T\mathbf{K}\mathbf{y} - C(\mathbf{G}, \mathbf{H}, \mathbf{K})\right\}, \qquad \mathbf{y} \in \mathbb{R}^m, \tag{17}$$

that was given in (6) is a normalizable density. Let $\mathbf{K}^+ = \big(\mathcal{I}\mathbf{K}\mathcal{I}\big)^-$ and rewrite (5) as

$$\begin{aligned} \log f(\mathbf{y}|\mathbf{V} = \mathbf{v}) &= \mathbf{v}^T\mathbf{H}\mathbf{y} - \frac{1}{2}\mathbf{y}^T\mathbf{K}\mathbf{y} \\ &= \mathbf{v}^T\mathbf{H}\mathbf{y} - \mathbf{v}^T\mathbf{H}\mathbf{K}^+\mathbf{H}\mathbf{v} + \mathbf{v}^T\mathbf{H}^T\mathbf{K}^+\mathbf{K}\mathbf{K}^+\mathbf{H}\mathbf{v} - \frac{1}{2}\mathbf{y}^T\mathbf{K}\mathbf{y} \end{aligned}$$

Using the notation from (7) and applying (16), the normalizing constant of the density in (1) is found to be given by

$$C(\mathbf{G}, \mathbf{H}, \mathbf{K}) = \log \sum_{\mathbf{v} \in \{0,1\}^m} \exp\left[\mathbf{v}^T\mathbf{G}\mathbf{v} + \mathbf{h}^T\big(\mathcal{I}\mathbf{K}\mathcal{I}\big)^-\mathbf{h}/2\right] \left[\det{}^+\left(\frac{1}{2\pi}\mathcal{I}\mathbf{K}\mathcal{I}\right)\right]^{1/2}.$$

9.4. *The anisometric penalty is a score test of $\theta_a = 0$ for all $a$.*

PROPOSITION 1. *Let* $\mathbf{H} = \left[\frac{\partial^2 \log f_{[b|A]}(\mathbf{y})}{\partial\theta_i\theta_j}\right]$ *be the conditional information. Suppose* $\mathbf{H}$ *and thus also its inverse* $\mathbf{H}^{-1}$ *is block-diagonal. Then the anisometric group lasso penalty is equivalent to a score test of the null hypothesis that* $\theta = 0$ *vs. the alternative that a pre-specified subvector* $\theta_a \neq 0$.

Proof: Let $c = \mathcal{V} \setminus \{a, b\}$ and suppose that $\theta_c = 0$. From the KKT conditions, $\theta_a = 0$ is an optimum if and only if

$$\nabla_a^T H_{aa}^{-1} \nabla_a < \lambda^2,$$

where $\nabla_a = \frac{\partial \log f_{[b|A]}(\mathbf{y})}{\partial\theta_a}$ is the $a$-subvector of the conditional log-likelihood gradient. Taking $\lambda^2$ to be an appropriate quantile from a $\chi^2$-distribution with $\dim(H_{aa})$ degrees of freedom results yields a score test.

9.5. *Simulation details.*

9.5.1. *Graphs and parametric alternatives.* In the *G-minimal* and *complete* scenarios, the underlying graph is a (perhaps incomplete) chain, with either 1.5% of nodes connected (Figure 4a) or 5% (Figures 4b and 5). In the *e. coli* scenario, the underlying graph is a 500-vertex subgraph sampled from a network described in Gama-Castro et al. [2011] and available from GeneNetWeaver [Schaffter et al., 2011]. In the $G$ parametric alternative, given the underlying graph, the data are derived from model (6), restated in (17), with only the $G$ interaction matrix set to non-zero. In this case, although the specified conditional independences hold exactly, an auto-logistic (Ising) model is minimally complete, while the multivariate Hurdle model is over-parametrized. In the *complete* parametric alternative, given the underlying graph, all three interaction matrices $G$, $H$ and $K$ are non-zero simultaneously in the appropriate entries.

9.5.2. *Generative models.* The Hurdle generative model, and deviations from it are considered. In the *exact* case, observations are generated through Gibbs sampling from model (6) using the full conditional distributions available in (8). Samples from conditional distributions are generated simply as Bernoulli and Normal random variates. A 2000 iteration burn-in phase, and sample thinning was employed. Thinned samples exhibited only mild auto-correlation. In the *contaminated* case, a matrix of exact variates $\mathbf{Y}$ are sampled, and onto them (given $Y_{ij} \neq 0$) is added $t_8$-distributed noise. So the final variates remain zero-inflated, but are heavier-tailed than a Normal distribution. In the *selection* case, a matrix $\tilde{\mathbf{Y}}$ of latent, non-zero-inflated

Gaussian variates are sampled that follow the graphical model implied by the $K$-interaction matrix. These are zero-inflated through a selection model

$$P\left(\tilde{V}_j | \mathbf{Y} = \mathbf{y}\right) = \text{logit}(a_j + b_j y_j),$$
$$\mathbf{Y} = \tilde{\mathbf{Y}}\tilde{V}.$$

The parameters $a_j$ and $b_j$ are chosen to keep $P(\tilde{V_j})$ away from the boundary values 0 and 1.

Lastly, in some cases, we consider *in-silico 10-cell* replicates. Given a desired sample size $n$, draw $10n$ observations $\mathbf{Y}$ from model (6), and let the observed data $\mathbf{Y}^{(\mathbf{10})}$ follow

$$\mathbf{Y}^{(\mathbf{10})} = \log_2 \sum_{i=1}^{10} 2^{\mathbf{Y}_i}/10.$$

**Acknowledgments.** RG was funded by a grant from the Bill and Melinda Gates foundation, the Vaccine and Immunology Statistical Center (VISC), OPP1151646. AM and RG acknowledge funding through grant R01 EB008400 from the National Institute of Biomedical Imaging and Bioengineering, US National Institutes of Health. MD was partially supported by grant DMS 1561814 from the US National Science Foundation.

AM thanks Daniel Lu for comments on the networks in reported section 6.

## References.

E-mail: andrew_mcdavid@urmc.rochester.edu

E-mail: rgottard@fredhutch.org

E-mail: nrsimon@uw.edu

E-mail: md5@uw.edu

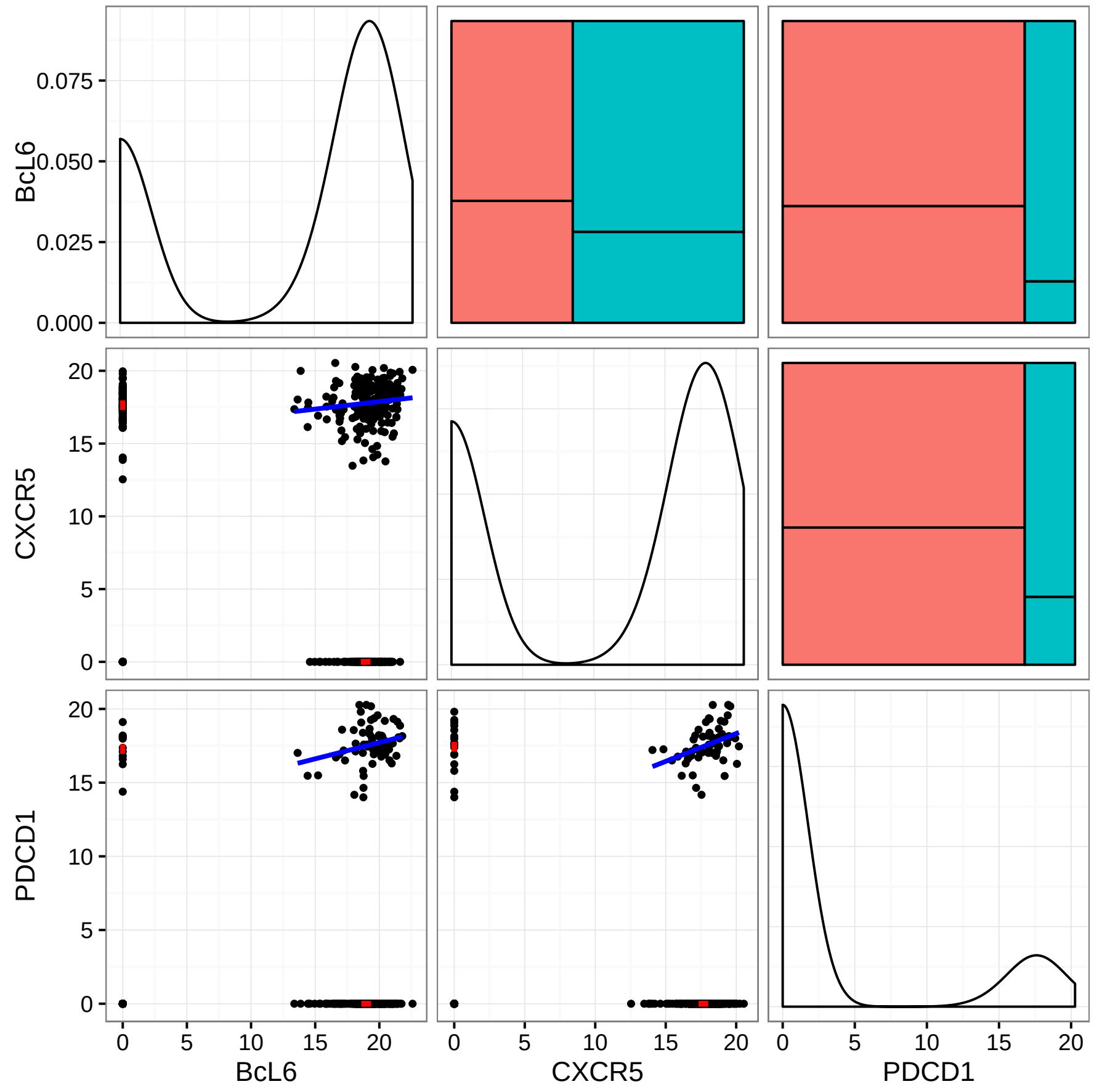

Fig 2: Scatter plots of inverse cycle threshold (40-Ct) measurements $\mathbf{y}$ from a quantitative PCR (qPCR)-based single cell gene expression experiment (lower panels). The cycle threshold (Ct) is the PCR cycle at which a pre-defined fluorescence threshold is crossed, so a larger inverse cycle threshold corresponds to greater log-expression [McDavid et al., 2013]. Measurements that failed to cross the threshold after 40 cycles are coded as 0.

The upper panels show mosaic plots of each pair of contigency tables that can be formed from the indicator functions $[\mathbf{v_y}]_i = I_{\{y_i \neq 0\}}$. On the lower panels, the linear regression on positive pairs of observations is indicated in blue, while the conditional mean values are indicated in red.

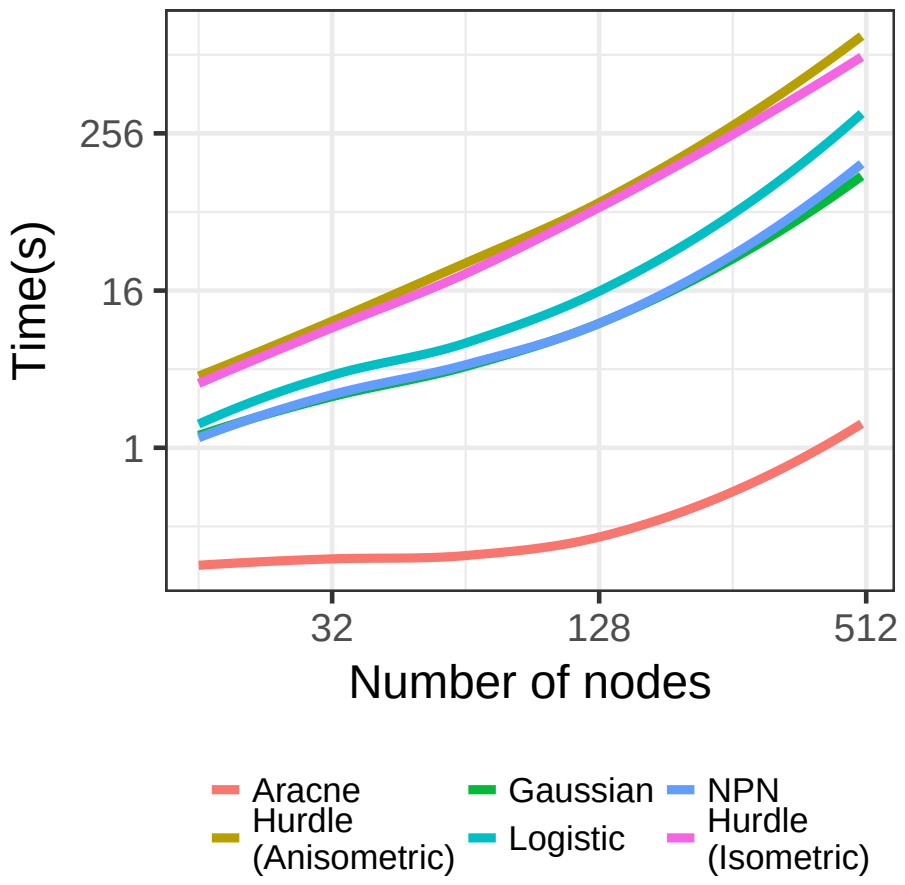

Fig 3: Average timings for graph estimation algorithms as a function of the number of nodes

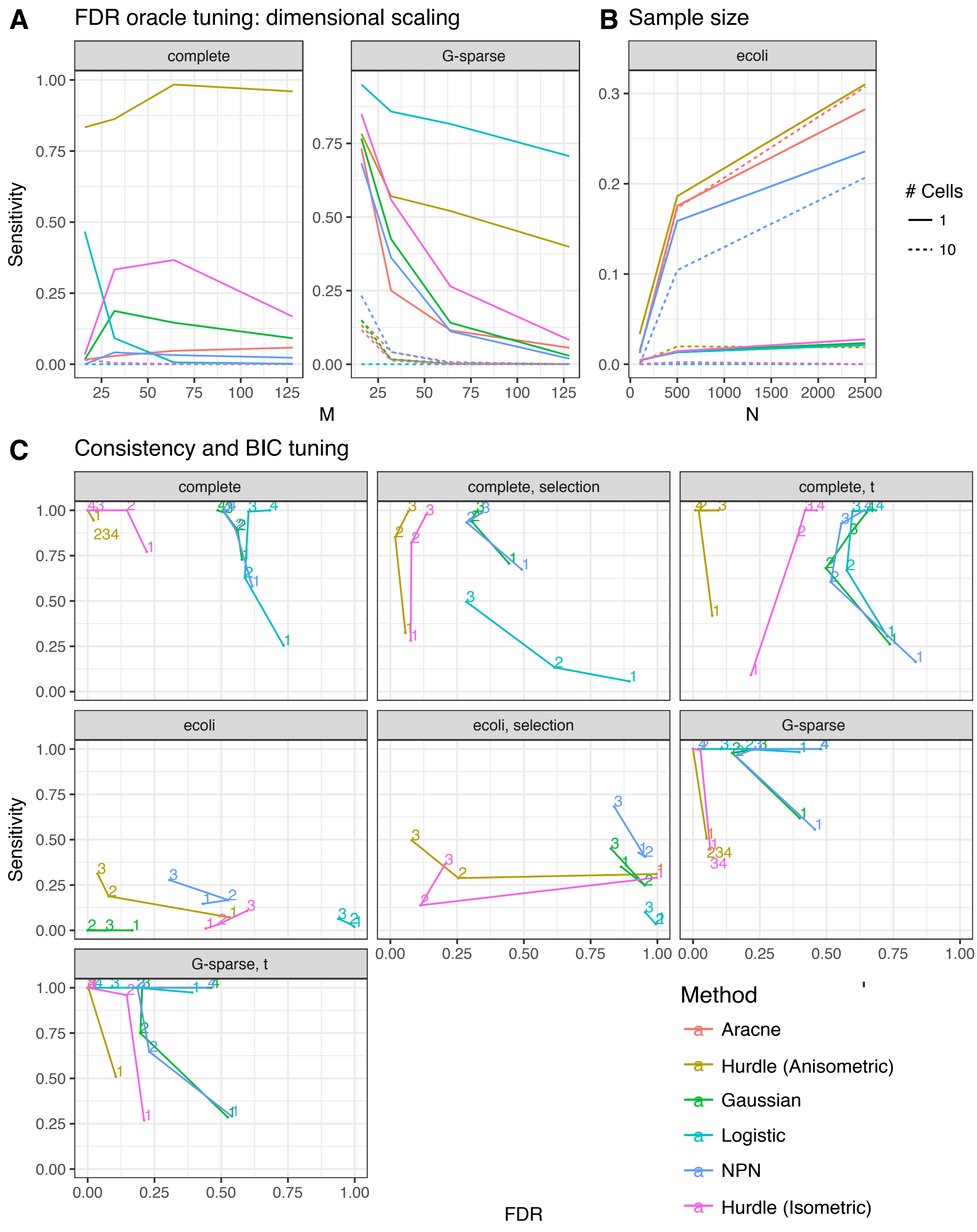

Fig 4: Dimensional (a) and sample size scaling (b) of six different network inference algorithms applied to simulated data under oracle FDR tuning. Data are generated from the multivariate hurdle model (6) under chain graphs (a) and *e. coli* graph (b). Panel (c) shows network selection consistency of various methods using the Bayesian Information Criterion under various models described in table 1. The paths trace out the changes in FDR and sensitivity as the sample size increases geometrically from 100 (1) to 12,500 (4).

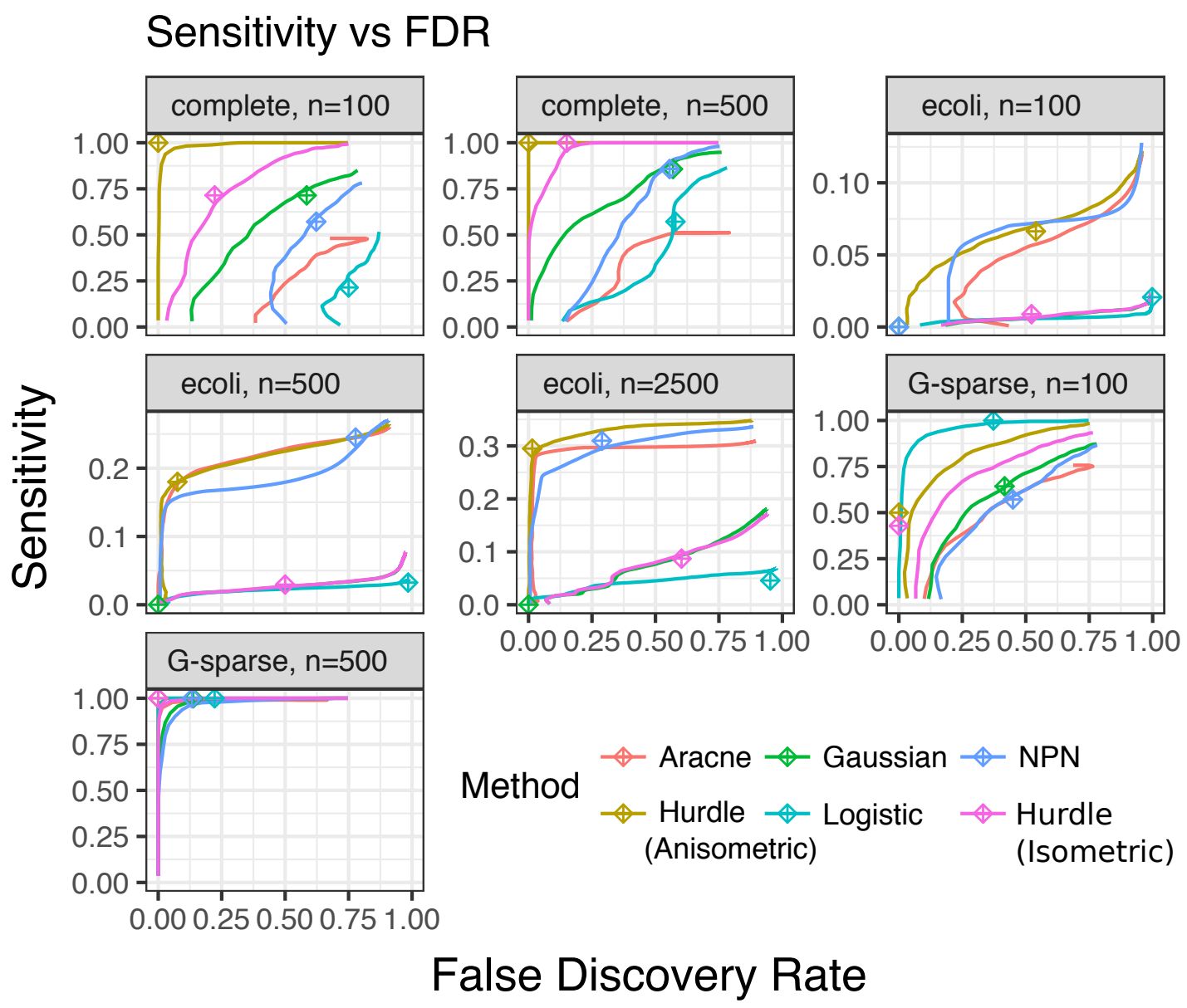

Fig 5: Sensitivity vs. FDR for solution paths from methods and scenarios described in Table 1. The ⊕ symbol indicates the tuning selected by BIC.

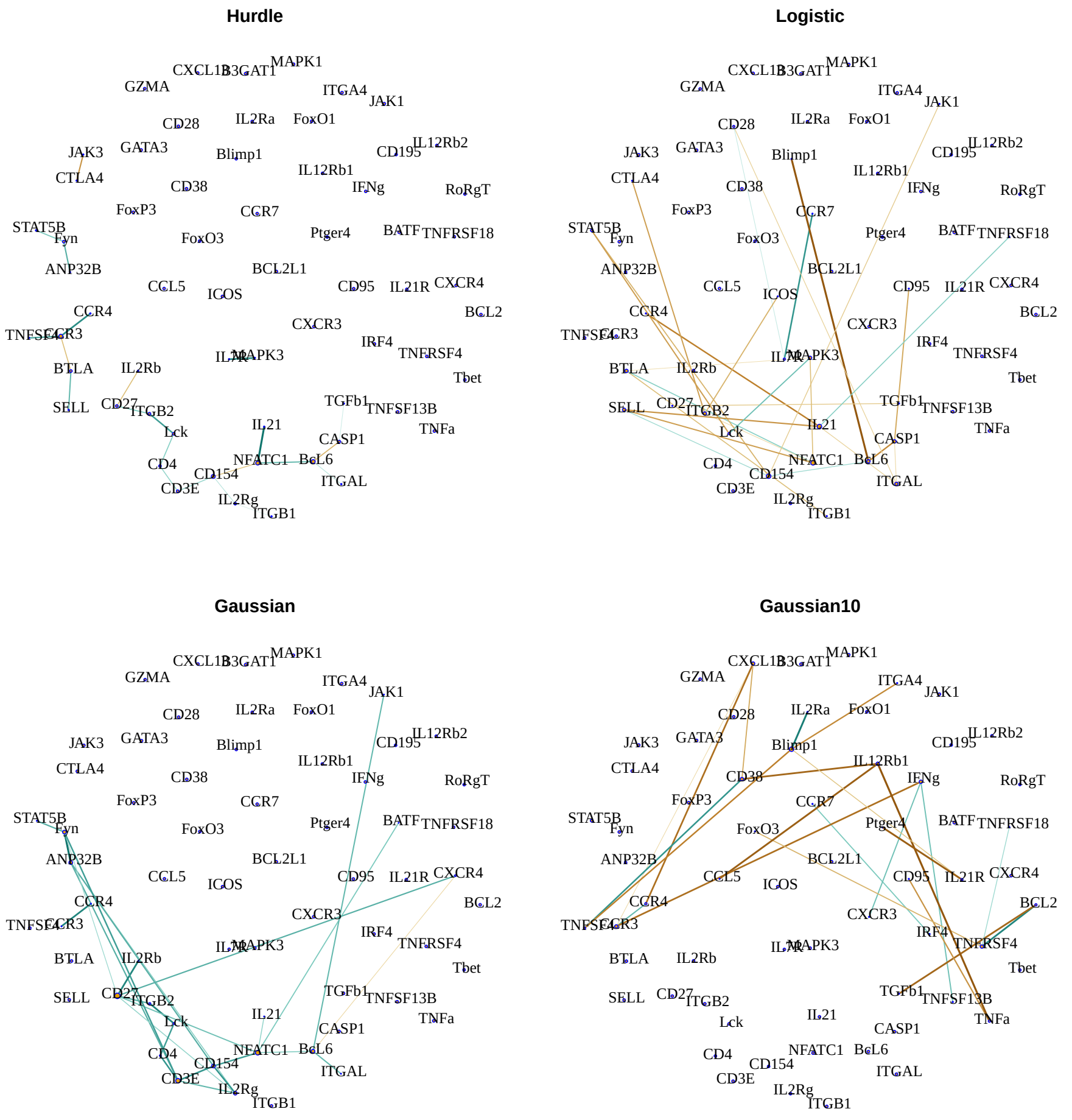

Fig 6: Networks of 22 edges estimated through neighborhood selection under the Aracne, Hurdle, logistic, Gaussian model (single cells) and Gaussian model (10 cell aggregates) in T follicular helper cells. Brown hues indicate estimated negative dependences, while blue-green hues indicate positive dependences. The edge width and saturation are larger for stronger estimated dependences.

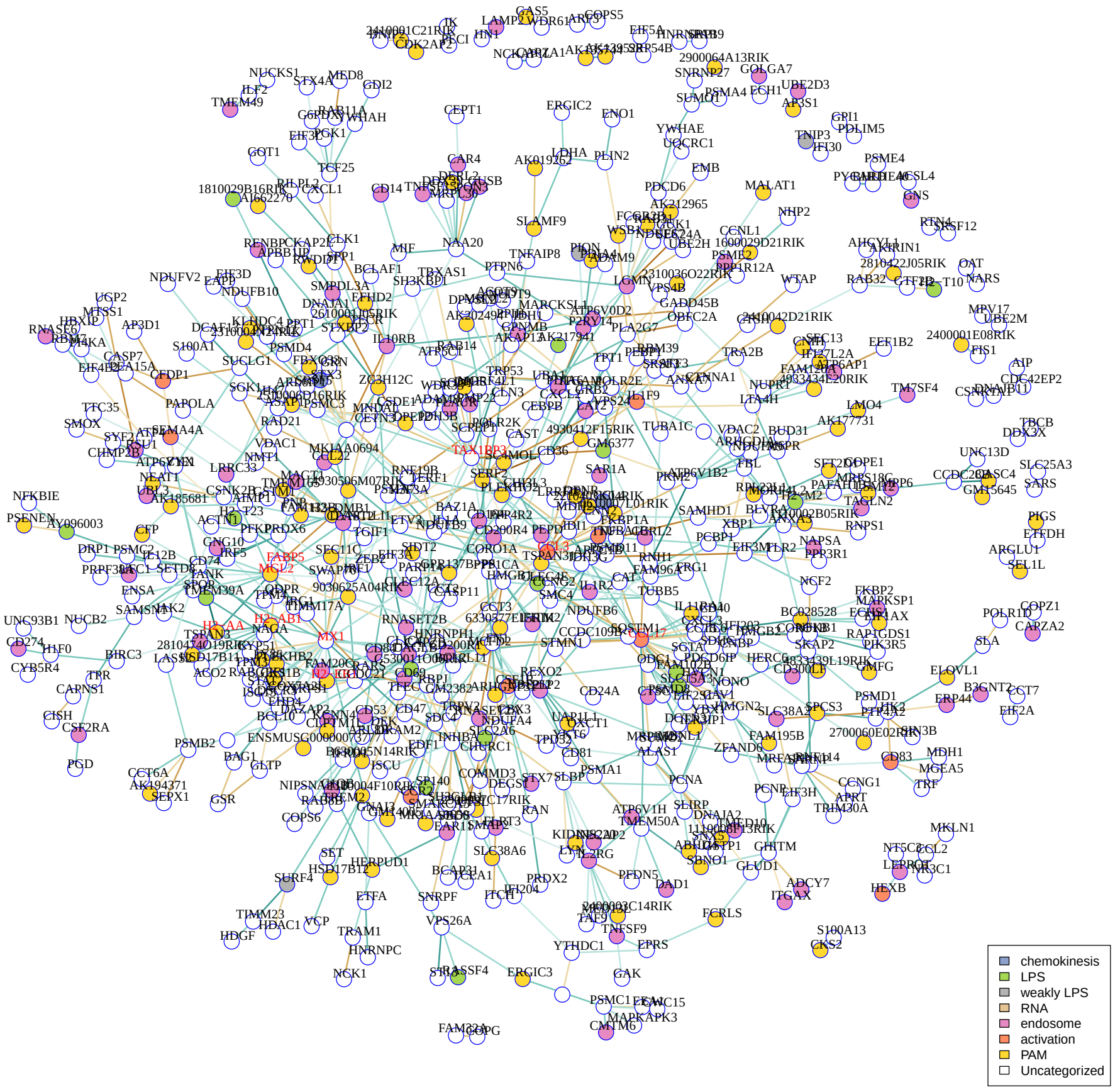

Fig 7: Core Gaussian model networks in LPS-treated mouse dendritic cells. Hub genes are shown in red. Vertex colors indicate gene ontology membership. Disconnected subgraphs with two vertices are suppressed.

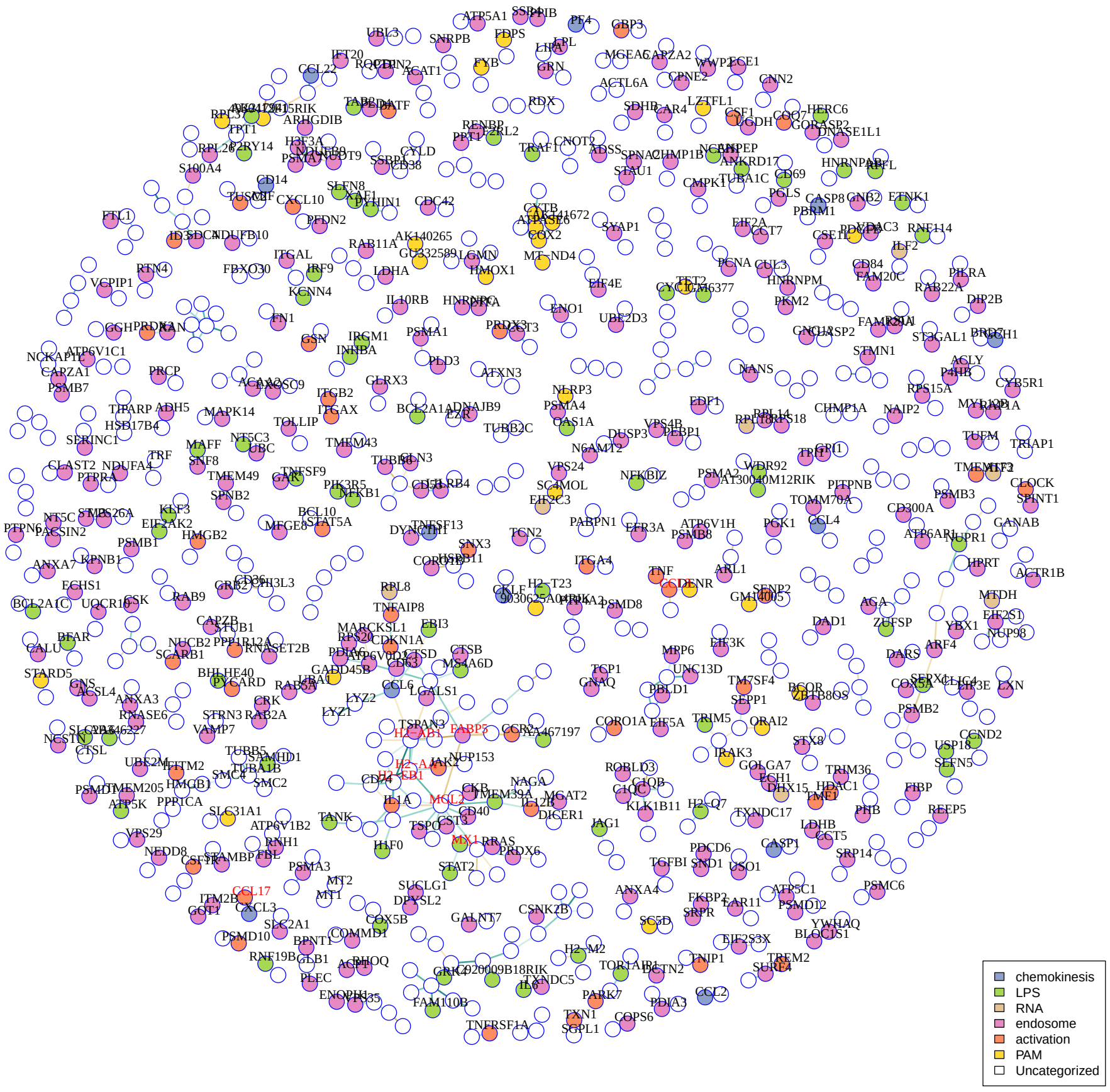

Fig 8: Core Hurdle model networks estimated in LPS-treated mouse dendritic cells. Hub genes are shown in red. Vertex colors indicate gene ontology membership. Disconnected subgraphs with two vertices are suppressed.

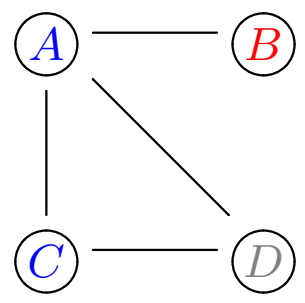

Fig 9: Overview of geneset edge enrichment analysis.

1. Vertices A and C belong to the blue category, while vertex B belongs to the red category. Vertex D belongs to neither.
2. There is $n_{ij} = 1$ blue-red intra-connection, while $m_{ij} = 2$ are possible given the 4 edges.
3. The enrichment statistic is the hypergeometric tail probability $t_{ij} = P(N = 2; 4, 2, 6) = \frac{\binom{2}{2}\binom{4}{2}}{\binom{6}{4}} = .4$
4. The significance of the blue-red enrichment statistic would be ascertained by sampling from the null Erdos-Renyi model over all possible pairs of categories.

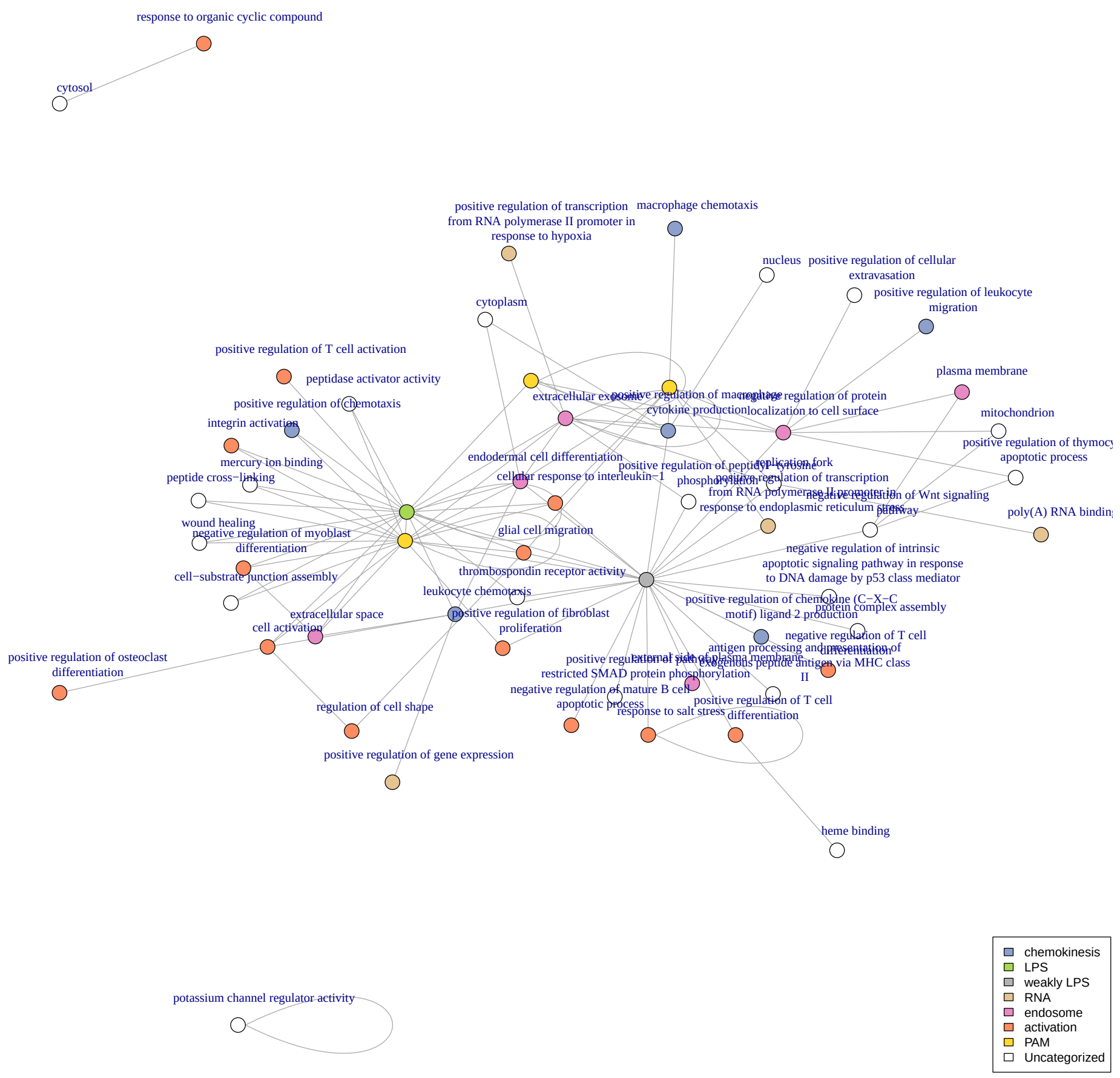

Fig 10: Modules enriched at FDR $\leq$ 10% using graphical geneset edge enrichment in mouse dendritic cells under Gaussian model.

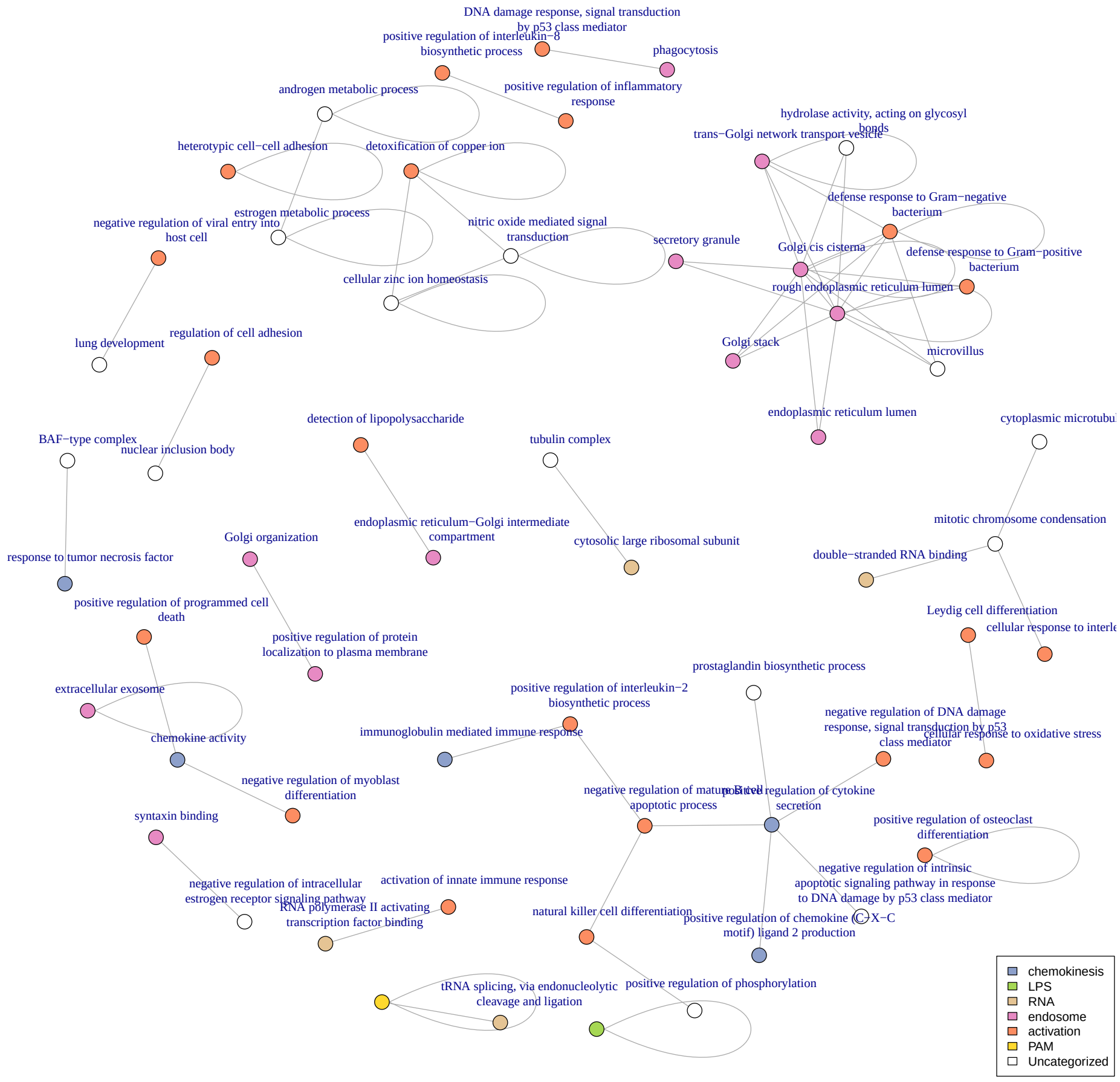

Fig 11: Modules enriched at FDR $\leq 10\%$ using graphical geneset edge enrichment in mouse dendritic cells under Hurdle model.

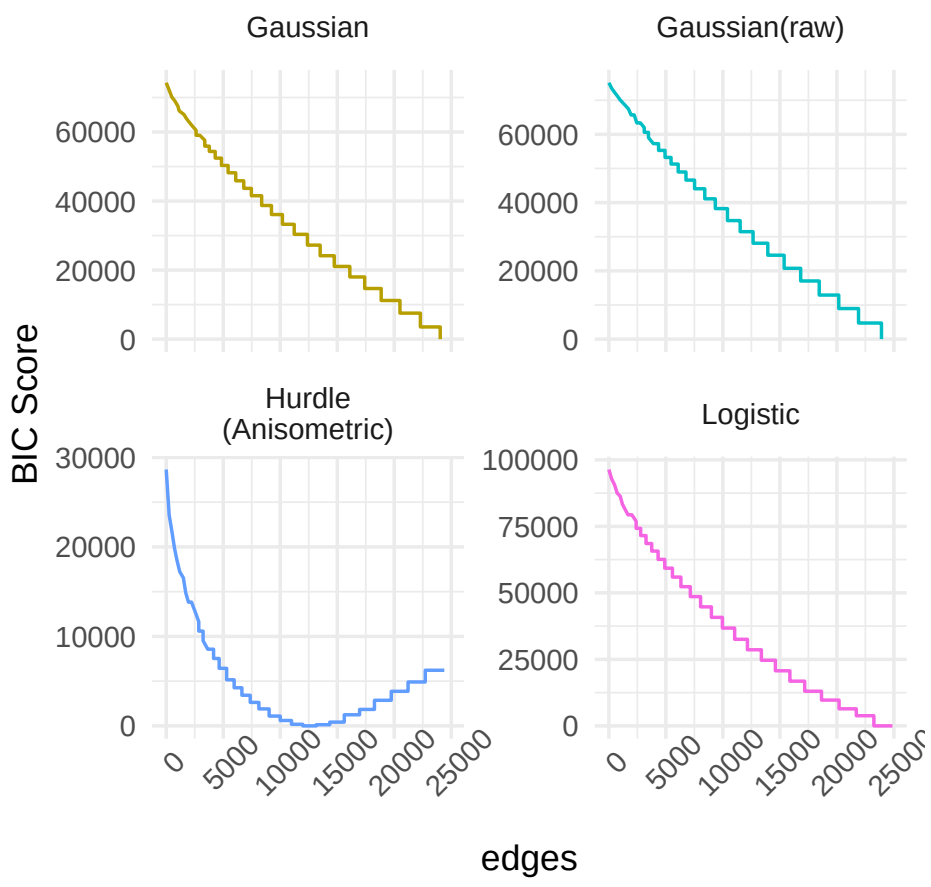

Fig 12: Bayesian information criterion on mDC data set